\documentclass[showpacs,preprintnumbers,amsmath,amssymb]{revtex4}

\usepackage{graphicx}% Include figure files
\usepackage{dcolumn}% Align table columns on decimal point
\usepackage{bm}% bold math
\usepackage{url}
\usepackage{epsfig}
\usepackage{multirow}
\usepackage{tensor}
\usepackage{empheq}
\usepackage{amsmath}
\usepackage{color}
\definecolor{myblue}{rgb}{.8, .8, 1}

\def\be{\begin{equation}}
\def\ee{\end{equation}}
\def\ba{\begin{eqnarray}}
\def\ea{\end{eqnarray}}

\newcommand{\fr}[2]{\frac{#1}{#2}}

\def\m{\rm{m}}

\def\D{\rm{D}}
\def\EdS{\rm{EdS}}

\newcommand{\p}{\textbf{p}}
\newcommand{\q}{\textbf{q}}
\newcommand{\s}{\textbf{S}}
\newcommand{\x}{\textbf{x}}

\newcommand{\kk}{\textbf{k}}

\newcommand{\PRD}{Phys.\ Rev.\ D}

\newcommand{\JCAP}{J.\ Cosmol.\ Astropart.\ Phys.}
\newcommand{\MNRAS}{Mon.\ Not.\ Roy.\ Astron.\ Soc.}

\newcommand{\DE}{\rm{DE}}
\def\ga{\mathrel{\raise.3ex\hbox{$>$\kern-.75em\lower1ex\hbox{$\sim$}}}}
\def\la{\mathrel{\raise.3ex\hbox{$<$\kern-.75em\lower1ex\hbox{$\sim$}}}}

\begin{document}

\title{Lagrangian Perturbation Theory : Exact One-Loop Power Spectrum in General Dark Energy Models}

%\\

\author{Seokcheon Lee}
\email[]{skylee@kias.re.kr}
\affiliation{School of Physics, Korea Institute for Advanced Study, Heogiro 85, Seoul 130-722, Korea}

\leftline{KIAS-P14021}

%\date{May 11, 2010}% It is always \today, today,

\begin{abstract}

Recently, we find that the correction for the EdS assumption on the one-loop matter power spectrum for general dark energy models using the standard perturbation theory is not negligible \cite{LPB}. Thus, we investigate the same problem by obtaining the exact displacement vector and kernels up to the third order for the general dark energy models in the Lagrangian perturbation theory (LPT). Using these exact solutions, we investigate the present one-loop matter power spectrum in the $\Lambda$CDM model with $\Omega_{m0} = 0.25 (0.3)$ to obtain about 0.2 (0.18) \% error correction compared to that obtained from the EdS assumption for $k = 0.1 \rm{h\, Mpc}^{-1}$ mode. If we consider the total matter power spectrum, the correction is only 0.05 (0.03) \% for the same mode. It means that EdS assumption is a good approximation for $\Lambda$CDM model in LPT theory. However, one can use this method for general models where EdS assumption is improper.

\end{abstract}


\maketitle

%%%%%%%%%%%%%%%%%%%%%%%%%%%%%%%%%%%%%%%%%%%%%%%%%%%%%%%%
%\setcounter{equation}{0}
%%%%%%%%%%%%%%%%%%%%%%%%%%%%%%%%%%%%%%%%%%%%%%%%%%%%%%%%
With the upcoming precision measurements of the large scale structure, the accurate theoretical modeling is essential to interpret the observational data. It requires huge number of mock catalogs and N-body simulations are too numerically expensive to be done. Fortunately, it seems that observable quantities at the quasi-linear scales might be accurately modeled semi-analytically. The Lagrangian perturbation theory (LPT) has been widely used to investigate for this purpose \cite{07112521,08071733,Err,12090780,13061804}. Also, the initial condition for N-body simulation are generated using LPT \cite{9711187,0606505,13092243}.

In LPT, the fundamental object is the Lagrangian displacement vector $\textbf{S}$, which displaces the particle from its initial position $\q$ to the final Eulerian position $\x$
\be \x(\q,t) = \q + \textbf{S}(\q,t) \, . \label{vecx} \ee
The first order LPT solution is the Zel'dovich approximation \cite{ZA} and higher order solutions have been obtained \cite{9309055,9406013,9406016,12034260,14012226}. From the mass conservation, the matter density perturbation $\delta$ can be described as a function of the $\s$
\be \delta (\x,t) = \int d^3 q \delta_{\D}\Bigl(\x - \q - \s(\q,t)\Bigr) - 1 \, . \label{deltax} \ee
One can expand the displacement vector $\s$ according to the Lagrangian perturbative prescription
\ba \s(\q,t) &\equiv& \sum_{n=1} \s^{(n)}(\q,t) = \sum_{n=1} D_{(n)}(t) \s^{(n)}(\q) = \sum_{n=1} I_{n}(t) D^{n}(t) \s^{(n)}(\q) \nonumber \\
&\equiv& D(t) \s^{(1)}(\q) + E(t) \s^{(2)}(\q) + F_{a}(t) \s^{(3a)}(\q) +
F_{b}(t) \s^{(3b)}(\q) + \cdots \, . \label{vecSn} \ea This explicit separation with respect to the spatial and temporal coordinates for each order ({\it i.e}. $I_{n}$ is a constant) is known to be a property of the perturbative Lagrangian description for an Einstein-de Sitter (EdS) universe \cite{9406016}. However, the solution at each order can be a separable function of $t$ and $\q$ even for general dark energy models by using $D_{(n)}(t)$ instead of $D^{(n)}(t)$. After one includes the time dependent of $I_{n}$ in the each kernel, one can find the exact solution for each order. One can use $I_{1} = 1$, $D_{1} = D$ where $D_1(t)$ is the linear growth factor, and $D_{(n)}(t) = I_{n}(t)D^{n}(t)$ are specified as,
\ba D_{(2)}(t) &\equiv& E(t) = I_{2}(t) D^{2}(t) \, , \label{Et} \\
D_{(3a)}(t) &\equiv& F_{a}(t) = I_{3a}(t) D^{3}(t) \, , \label{Fat} \\
D_{(3b)}(t) &\equiv& F_{b}(t) = I_{3b}(t) D^{3}(t) \, . \label{Fbt} \ea
From these equations (\ref{Et})-(\ref{Fbt}), one can obtain Lagrangian Poisson equation order by order (from the linear to the irrotational third orders)
\ba \ddot{D} + 2H \dot{D} - 4 \pi G \rho_{m} D &=& 0 \, , \label{Deq} \\
\ddot{E} + 2H \dot{E} - 4 \pi G \rho_{m} E &=& - 4 \pi G \rho_{m}  D^2 \, , \hspace{0.1in}  {\rm if} \,\,\, \mu_{1}(\s^{(2)} ) = \mu_{2}(\s^{(1)},\s^{(1)}) \, , \label{Eeq} \\
\ddot{F}_{a} + 2H \dot{F}_{a} - 4 \pi G \rho_{m} F_{a} &=& - 8 \pi G \rho_{m} D^3 \, , \hspace{0.1in} {\rm if} \,\,\, \mu_{1}(\s^{(3a)} ) = \mu_{3}(\s^{(1)}) \, , \label{Faeq} \\
\ddot{F}_{b} + 2H \dot{F}_{b} - 4 \pi G \rho_{m} F_{b} &=& - 8 \pi G \rho_{m} D ( E - D^2 ) \, , \hspace{0.1in} {\rm if} \,\,\, \mu_{1}(\s^{(3b)}) = \mu_{2}(\s^{(1)},\s^{(2)}) \, , \label{Fbeq} \ea
where dots represent the derivatives with respect to the cosmic time $t$ and $\mu_{2}(S^{(1)},S^{(2)}) = \mu_{2}(S^{(2)},S^{(1)})$ is satisfied for any tensor \cite{12034260}. $\mu_{a}(\s^{(n)})$ are defined as
\ba \mu_{1}(\s^{(n)}) &\equiv& S_{ii}^{(n)} \, , \label{mu1} \\
\mu_{2}(\s^{(n)},\s^{(m)}) &\equiv& \fr{1}{2} \Bigl( S_{ii}^{(n)} S_{jj}^{(m)} - S_{ij}^{(n)} S_{ji}^{(m)} \Bigr) \, , \label{mu2} \\
\mu_{3}(\s^{(n)}) &\equiv& {\rm det} S_{ij}^{(n)} \, . \label{mu3} \ea
One can rewrite the above Eq. (\ref{vecSn}) in Fourier space represented by using the linear matter density contrast, $\widetilde{\delta}_{L}(p)$
\ba \widetilde{\s}^{(n)}(\kk,t) &=& - i D_{(n)}(t) \int \fr{d^3 p_1}{(2 \pi)^3} \cdots \fr{d^3 p_n}{(2 \pi)^3} (2\pi)^3 \delta_{\D}(\p_{1 \cdots n} - \kk) \textbf{F}^{(n)}(\p_1, \cdots, \p_n) \widetilde{\delta}_{L} (p_1) \cdots \widetilde{\delta}_{L} (p_n) \nonumber \\
&=& - i \fr{D^{n}(t)}{n !} \int \fr{d^3 p_1}{(2 \pi)^3} \cdots \fr{d^3 p_n}{(2 \pi)^3} (2\pi)^3 \delta_{\D}(\p_{1 \cdots n} - \kk) n ! I_{n}(t) \textbf{F}^{(n)}(\p_1, \cdots, \p_n) \widetilde{\delta}_{L} (p_1) \cdots \widetilde{\delta}_{L} (p_n) \nonumber \\
&\equiv& - i \fr{D^{n}(t)}{n !} \int \fr{d^3 p_1}{(2 \pi)^3} \cdots \fr{d^3 p_n}{(2 \pi)^3} (2\pi)^3 \delta_{\D}(\p_{1 \cdots n} - \kk) \L^{(n)} (t, \p_1, \cdots, \p_n) \widetilde{\delta}_{L} (p_1) \cdots \widetilde{\delta}_{L} (p_n) \label{tildeSn} \, . \ea
In the second equality, we adopt the same notation as in \cite{07112521,08071733,Err}. We also use $\p_{1 \cdots n} = \p_1 + \cdots + \p_n$, $\textbf{F}^{(n)} = (-1)^{n} \fr{\p_{1 \cdots n}}{p_{ 1\cdots n}^2} \fr{\kappa^{(n)}(\p_1, \cdots, \p_n)}{p_1^2 \cdots p_{n}^2} $, and the n-th order kernels $\kappa^{(n)}$ are same as given in \cite{9406016,12034260}. $I_{n}(a)$ should be obtained numerically from Eqs.(\ref{Eeq})-(\ref{Fbeq}) by using the EdS initial conditions given by Eqs.(\ref{EgEdS})-(\ref{FbgEdS}). In the literatures, one uses the coefficients for the EdS solutions
\be I_2 = -\fr{3}{7}, \, I_{3a} = -\fr{1}{3}, \, I_{3b} = \fr{10}{21}\, . \label{IEdSs} \ee
However, these values are the approximate ones with using EdS assumption and we will use the exact values of them.

From the above equation (\ref{tildeSn}), one can obtain the perturbative kernels in LPT up to the third order
\ba L^{(1)}(\p_1) &=& \fr{\kk}{k^2} \, , \label{L1} \\
L^{(2)}(a,\p_1,\p_2) &=& I_2(a) \fr{\kk}{k^2} \Biggl[1 - \Bigl(\fr{\p_1 \cdot \p_2}{p_1p_2} \Bigr)^2 \Biggr] \, , \label{L2} \\
L^{(3a)}(a,\p_1,\p_2,\p_3) &=& I_{3a}(a) \fr{\kk}{k^2} \Biggl[1 - 3 \Bigl(\fr{\p_1 \cdot \p_2}{p_1p_2} \Bigr)^2 + 2 \fr{(\p_1\cdot\p_2)(\p_2\cdot\p_3)(\p_3\cdot\p_1)}{p_1^2\,p_2^2\,p_3^2} \Biggr] \, , \label{L3a} \\
L^{(3b)}(a,\p_1,\p_2,\p_3) &=& I_{3b}(a) \fr{\kk}{k^2} \Biggl[1 - \Bigl(\fr{\p_1 \cdot \p_2}{p_1p_2} \Bigr)^2 \Biggr] \Biggl[1 - \Bigl(\fr{(\p_1+\p_2) \cdot \p_3}{|p_1+p_2|p_3} \Bigr)^2 \Biggr] \, . \label{L3b} \ea
The above kernels are identical to those of EdS given in \cite{07112521,12034260} when $I_{2}$-$I_{3b}$ are given by Eq.(\ref{IEdSs}).

From the above consideration, one can obtain the non-linear power spectrum with one-loop correction by using a resummation scheme known as integrated perturbation theory \cite{07112521,08071733}
\ba P(k) &=& \exp \Biggl[-k_ik_j \int \fr{d^3 p}{(2\pi)^3} C_{ij}(\p) \Biggr] \Biggl[ k_ik_j C_{ij}(\kk) + k_ik_jk_k \int \fr{d^3 p}{(2\pi)^3} C_{ijk} (\kk,-\p,\p-\kk) \nonumber \\ &+& \fr{1}{2} k_ik_jk_kk_l \int \fr{d^3 p}{(2\pi)^3} C_{ij}(\p) C_{kl}(\kk-\p) \Biggr] \label{Pk} \, , \ea
where the mixed polyspectra of linear density field and the displacement field is defined as
\be \Bigl \langle \widetilde{\delta}_{L}(\kk_1)\cdots\widetilde{\delta}_{L}(\kk_l)\widetilde{S}_{i_1}(\p_1)\cdots\widetilde{S}_{i_m}(\p_m) \Bigr \rangle_{c} = (2\pi)^3\delta_{\D}(\kk_{1 \cdots l}+\p_{1 \cdots m})(- \imath)^m C_{i_i\cdots i_m}(\kk_1,\cdots,\kk_l;\p_1,\cdots,\p_m) \label{Cij1} \, . \ee
Also the mixed polyspectra of each order in perturbations is given by
\be \Bigl \langle \widetilde{\delta}_{L}(\kk_1)\cdots\widetilde{\delta}_{L}(\kk_l)\widetilde{S}^{(n_1)}_{i_1}(\p_1)\cdots\widetilde{S}^{(n_m)}_{i_m}(\p_m) \Bigr \rangle_{c} = (2\pi)^3\delta_{\D}(\kk_{1 \cdots l}+\p_{1 \cdots m})(- \imath)^m C^{(n_1\cdots n_m)}_{i_i\cdots i_m}(\kk_1,\cdots,\kk_l;\p_1,\cdots,\p_m) \label{Cij2} \, . \ee
From Eq. (\ref{Pk}) and (\ref{Cij2}), one can obtain the matter power spectrum with one-loop correction
\ba P(k) &=& \exp \Bigl[-k_i k_j \int \fr{d^3 p}{(2\pi)^3} C_{ij}^{(11)}(\p) \Bigr] \times \Biggl( k_i k_j
\Bigl[ C_{ij}^{(11)}(\kk) + C_{ij}^{(22)}(\kk) +C_{ij}^{(13)}(\kk) + C_{ij}^{(31)}(\kk) \Bigr] \nonumber \\
&+& k_i k_j k_k \int \fr{d^3 p}{(2\pi)^3} \Bigl[ C_{ijk}^{(112)}(\kk,-\p,\p-\kk)
+ C_{ijk}^{(121)}(\kk,-\p,\p-\kk) + C_{ijk}^{(211)}(\kk,-\p,\p-\kk) \Bigr] \nonumber \\
&+& \fr{1}{2} k_i k_j k_k k_l \int \fr{d^3 p}{(2\pi)^3} C_{ij}^{(11)}(\p) C_{ij}^{(11)}(\kk-\p) \Biggr) \label{Pkk}  \ea
After the analytic angular integration of the some of the Eq. (\ref{Pkk}), one obtains

\ba P(k) &=& \exp \Bigl[-\fr{k^2}{6 \pi^2} \int d p P_{L}(p) \Bigr] \nonumber \\ &\times& \Biggl[ P_{L}(k)
+ \fr{(2 \pi)^{-2} k^3}{2} \int_{0}^{\infty} dr P_{L}(kr) \int_{-1}^{1} dx P_{L} \Bigl( k\sqrt{1+r^2-2rx} \Bigr) \Biggl[ \fr{-I_2 r + x -(1 - I_2) r x^2}{(1 + r^2 - 2rx)} \Biggr]^2 \nonumber \\
&+& \fr{(2 \pi)^{-2} k^3}{48} P_{L}(k) \int_{0}^{\infty} dr P_{L}(kr)
\Biggl( -6(2I_2+I_{3b}) r^{-2} + 2(10I_2+11I_{3b}) + 2(-10I_2+11I_{3b}) r^2 \nonumber \\ &+& 6(2I_2-I_{3b}) r^4 + \fr{3}{r^3} (r^2-1)^3
\Bigl( (-2I_2+I_{3b}) r^2 - (2I_2+I_{3b}) \Bigr) \ln \Bigl|\fr{1+r}{1-r} \Bigr| \Biggr) \Biggr] \nonumber \\
&\equiv& \exp \Bigl[-\fr{k^2}{6 \pi^2} \int d p P_{L}(p) \Bigr]  \Biggl[ P_{L}(k) + P_{22}(k) + P_{13}(k) \Biggr] \equiv  \exp \Bigl[-\fr{k^2}{6 \pi^2} \int d p P_{L}(p) \Bigr]  \Bigl[ P_{NL}(k) \Bigr] \, , \label{Pkk2} \ea
where $r = \fr{p}{k}$ and $x = \fr{\p \cdot \kk}{p k}$. The above equations are identical to Eqs.(36) of \cite{07112521} when one replaces $I_2$ and $I_{3b}$ with those of EdS case given by Eq.(\ref{IEdSs}). Thus, the terms with $I_{2}$ and $I_{3b}$ represent the dark energy effect on the power spectrum. Also, both $I_{2}$ and $I_{3b}$ depend on the time and their values are changed depending on the measuring epoch. One interesting feature is that $I_{3a}$ does not contribute the one-loop correction in the matter power spectrum. When we generalize the power spectrum in the SPT without using the EdS assumption, we obtain the similar matter power spectrum as given in \cite{LPB}
\ba P^{\rm{SPT}}(k) &=& P_{L}(k) + \fr{(2 \pi)^{-2} k^3}{2} \int_{0}^{\infty} dr P_{L}(kr) \int_{-1}^{1} dx P_{L} \Bigl( k\sqrt{1+r^2-2rx} \Bigr) \Biggl[ \fr{(c_{21}+2c_{22}) r + (c_{21}-2c_{22}) x -2 c_{21} r x^2}{(1 + r^2 - 2rx)} \Biggr]^2 \nonumber \\
&+& (2\pi)^{-2} k^3 P_{L}(k) \int_{0}^{\infty} dr P_{L}(kr) \Biggl[ 2c_{35} r^{-2} -\fr{1}{3} \Bigl( 4 c_{31} -8 c_{32} +3c_{33} +24c_{35} - 16 c_{36} \Bigr) \nonumber \\ &-& \fr{1}{3}\Bigl(4 c_{31} -8c_{32} +12c_{33}-8c_{34}+6c_{35} \Bigr)r^2 + c_{33} r^4 + \Bigl(\fr{r^2-1}{r} \Bigr)^3 \ln \Bigl|\fr{1+r}{1-r} \Bigr| \Bigl(c_{35} - \fr{1}{2}c_{33}r^2 \Bigr) \Biggr] \label{PSPT} \, , \ea
where $c_{21}$-$c_{36}$ are also given in the above reference. If we adopt the EdS assumption both for the LPT matter power spectrum and the SPT one, then $P_{22}$ is same for the both approaches. However, the exact solutions will not be matched exactly for both cases. Also compared to SPT case where the magnitude of $P_{13}$ is comparable to that to $P_{13}$, the magnitude of $P_{13}$ is much smaller than that of $P_{22}$ in LPT. This will show that the EdS approximation in LPT is quite accurate.  

%%%%%%%%%%%%%%%%%%%%%%%%%%%%%%%%%%%%%%%%%%%%%%
\begin{figure}
\centering
\vspace{1.5cm}
\begin{tabular}{cc}
\epsfig{file=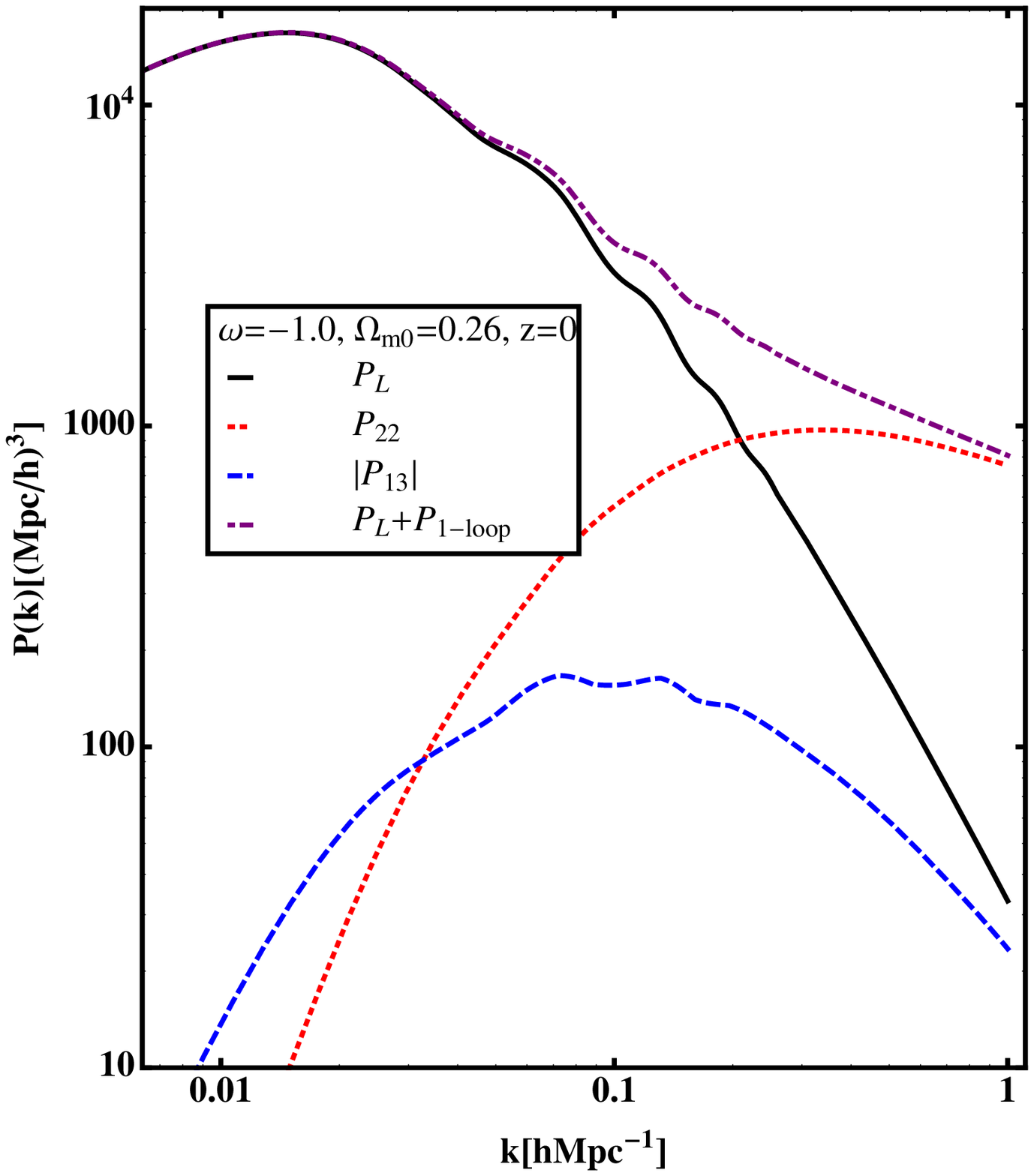,width=0.46\linewidth,clip=} &
\epsfig{file=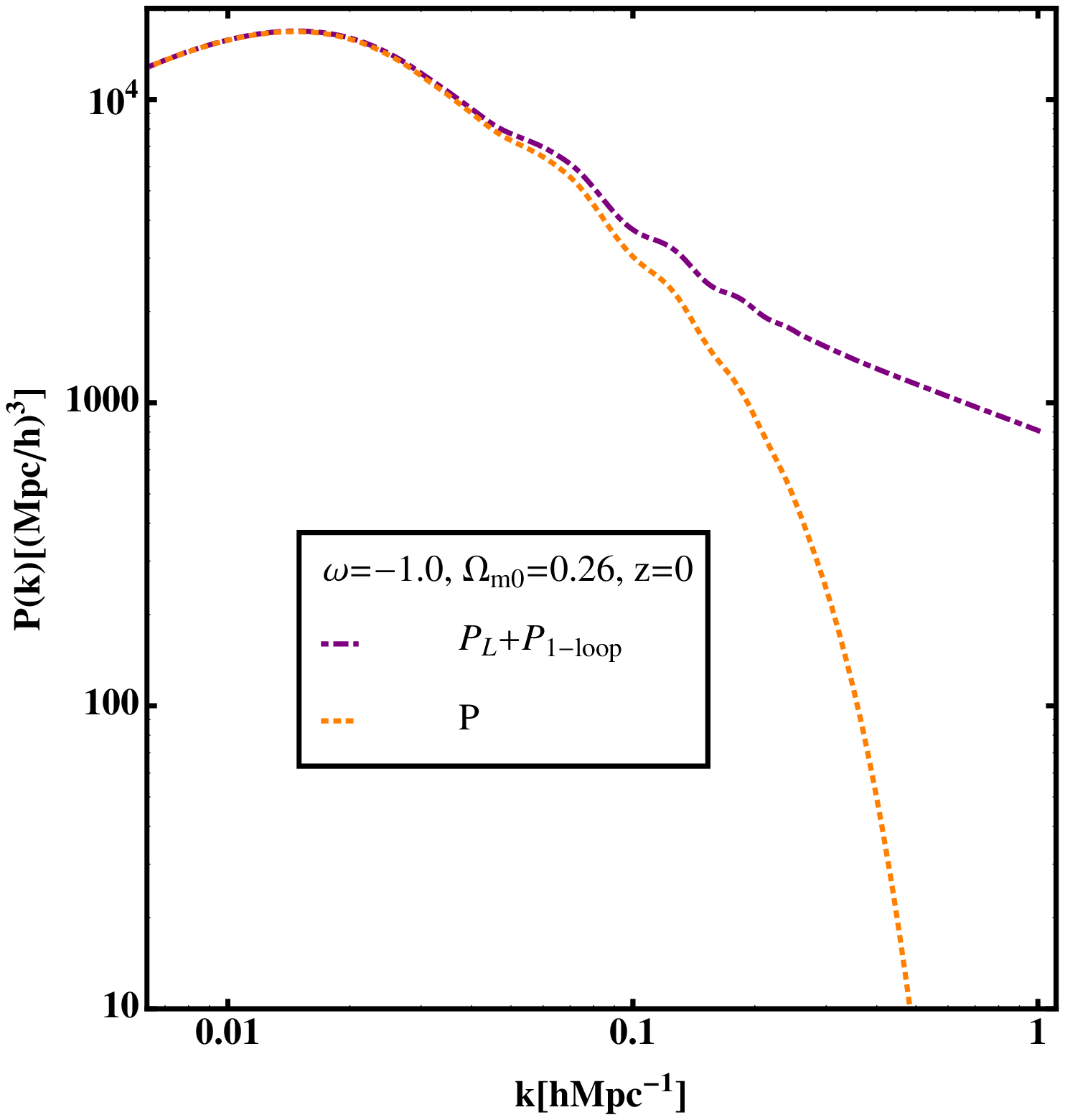,width=0.5\linewidth,clip=} \\
\end{tabular}
\vspace{-0.5cm}
\caption{$P_{NL}(k)$ and $P(k)$ a) Solid, dotted, dashed, and dot-dashed line represent $P_{L}$, $P_{22}$, $|P_{13}|$, and $P{NL}$, respectively. b) $P_{NL}$ and $P(k)$ are indicated as dot-dahed and dotted lines, respectively.} \label{fig1}
\end{figure}
%%%%%%%%%%%%%%

Now we obtain the one-loop power spectrum for $\Lambda$CDM model. We run the camb to obtain the linear power spectrum \cite{camb} using $\Omega_{b0} = 0.044$, $\Omega_{m0} = 0.26$, $h = 0.72$, $n_{s} = 0.96$, and the numerical integration range for $p$ in Eq. (\ref{Pkk2}) is $10^{-6} \leq p \leq 10^{2}$. In the left panel of Fig. \ref{fig1}, we show the linear power spectrum $P_{L}$ (solid), the one-loop power spectrum $P_{22}$ (dotted), $|P_{13}|$ (dashed), and the nonlinear power spectrum $P_{NL} = P_{L} + P_{1-\rm{loop}}$ (dot-dashed). Absolute magnitude of $P_{13}$ is smaller than that of SPT. Thus, the 1-loop correction is larger than that of SPT. 1-loop correction is mainly contributed from the $P_{22}$ and the coefficient $I_2$ is not much deviated from the that of EdS one ($-\fr{3}{7}$) as shown in the appendix. That is why the EdS approximation is a good one for LPT. Also, there exists the additional exponential prefactor to get the total power spectrum. This is shown in the right panel of Fig. \ref{fig1}. Dot-dashed line represents $P_{NL}$ and dotted line indicates $P(k)$.

%%%%%%%%%%%%%%%%%%%%%%%%%%%%%%%%%%%%%%%%%%%%%%
\begin{figure}
\centering
\vspace{1.5cm}
\begin{tabular}{cc}
\epsfig{file=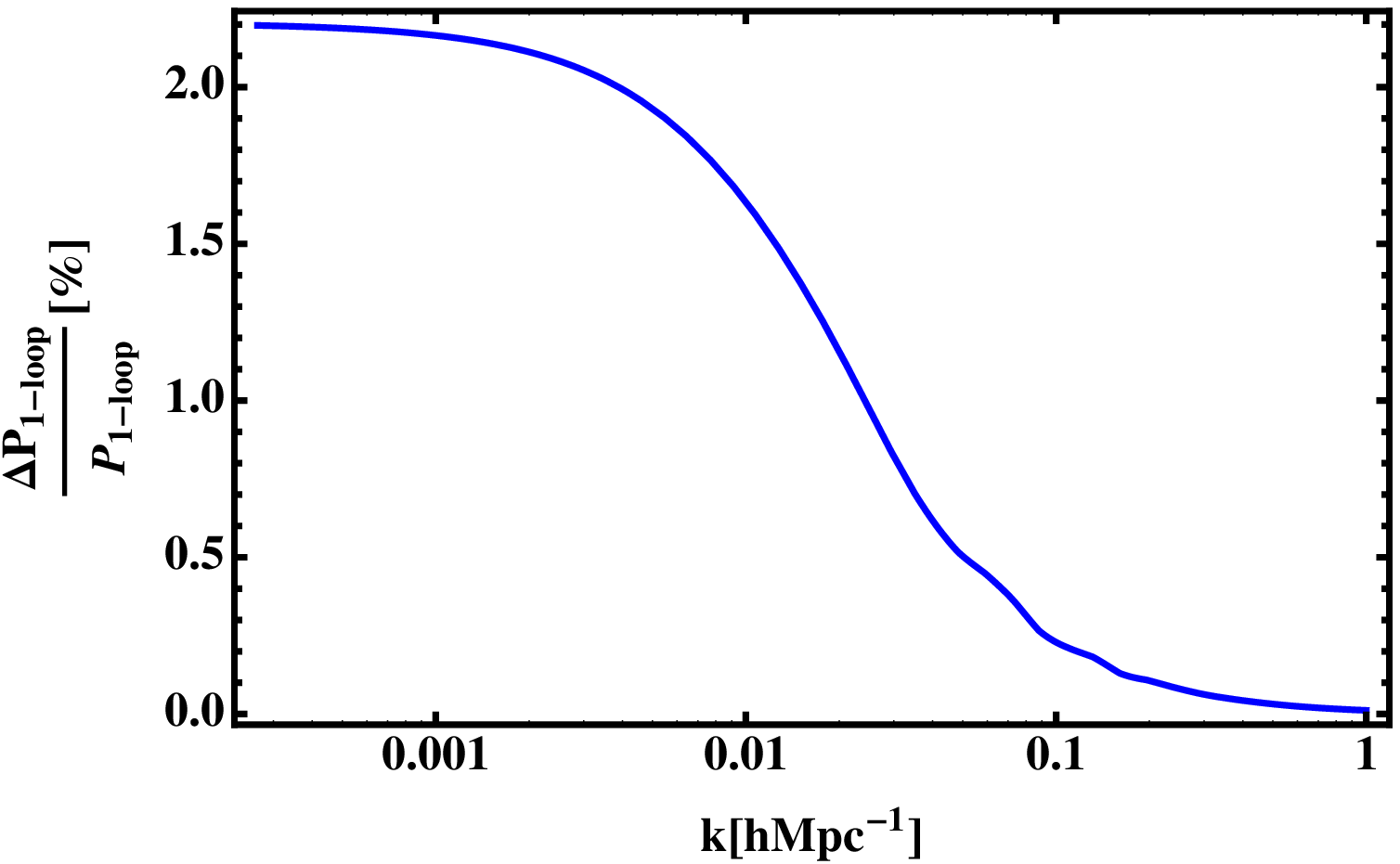,width=0.5\linewidth,clip=} &
\epsfig{file=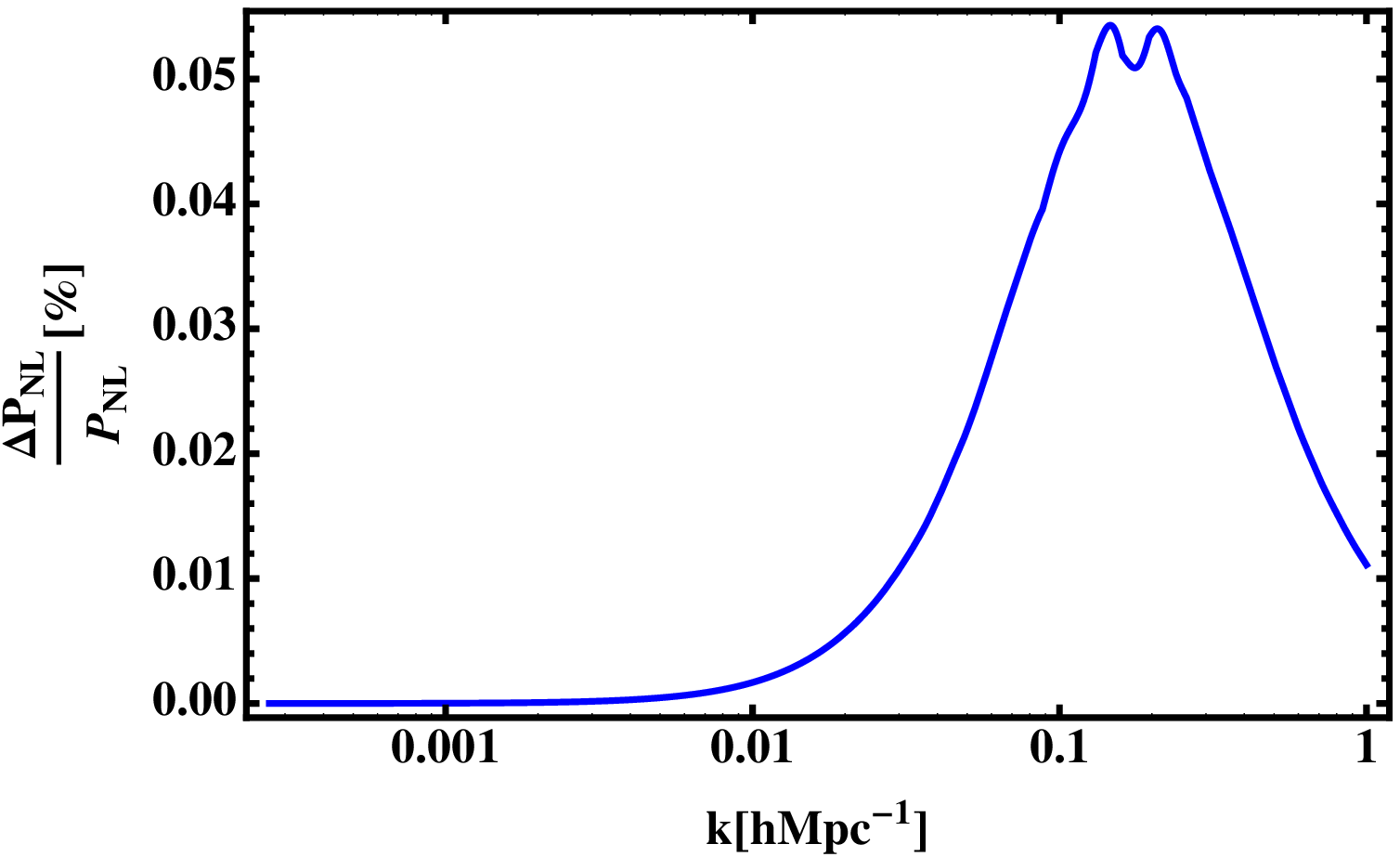,width=0.5\linewidth,clip=} \\
\end{tabular}
\vspace{-0.5cm}
\caption{Errors in $P_{1-\rm{loop}}$ and $P_{NL}$ a) The percentage difference between the correct $P_{1-\rm{loop}}$ and the one with EdS assumption. b) The percentage difference between $P_{NL}$ and $P_{NL}^{(\rm{EdS})}$.} \label{fig2}
\end{figure}
%%%%%%%%%%%%%%

Now we investigate the effect of dark energy on the 1-loop power spectrum compared to the one with EdS assumption. The difference in $P_{22} + P_{13}$ between them is shown in the left panel of Fig. \ref{fig2}. There exists only 0.2 \% error in $k = 0.1 \rm{h Mpc^{-1}}$ mode at present epoch. When we investigate them at the different $z$s, then the error is about the same. The error can be about 2 \% at large scales but 1-loop power spectrum is much smaller than the linear power spectrum at these scales. When we consider the total $P_{NL}$, the different is even smaller. The error is less than 0.05 \% for the same mode. This is shown in right panel of Fig. \ref{fig2} with notation $\Delta P_{NL} = P_{NL} - P_{NL}^{(\rm{EdS})}$. This proves the goodness of EdS assumption in LPT claimed in \cite{0112551}. However, we need to pay attention to this EdS assumption, when we consider more general models.

We show that EdS assumption is a good approximation to calculate the $\Lambda$CDM 1-loop power spectrum in Lagrangian perturbation theory. However, when we consider general dark energy models we need to consider the fully consistent method by using the fact $I_{n}$ depends on time. This also makes it possible to separate the temporal and spatial parts of solutions. We might be able to extend this method to the early dark energy or the modified gravity theories. The upcoming redshift surveys will provide observational data of large scale structure of the universe in larger volume with higher density. We obtain the accurate Lagrangian perturbation theory without using any assumption and this matches the requirement from future surveys. The obtained results are general for any background universe model including time varying dark energy models.

%%%%%%%%%%%%%%%%%%%%%%%%%%%%%%%%%%%%%%%%%%%%%%%%%%%%%%%%%%%%%%%%%%%%%%%%
\section*{Acknowledgments}
%%%%%%%%%%%%%%%%%%%%%%%%%%%%%%%%%%%%%%%%%%%%%%%%%%%%%%%%%%%%%%%%%%%%%%%%%
We would like to thank Sang-Gyu Biern, Xiao-Dong Li, and Cristiano Sabiu for useful discussion. This work were carried out using computing resources of KIAS Center for Advanced Computation. We also thank for the hospitality at APCTP during the program TRP.

%%%%%%%%%%%%%%%%%%%%%%%%%%%%%%%%%%%%%%%%%%%%%%%%%%%%%%%%%%%%%%%%%%%%%%%%
%\appendix
\renewcommand{\theequation}{A-\arabic{equation}}
% redefine the command that creates the equation no.
\setcounter{equation}{0}  % reset counter
\section*{APPENDIX}  % use *-form to suppress n1umbering
%*\section*{Appendix}
%\numberwithin{equation}{section}
%\setcounter{equation}{0}
%%%%%%%%%%%%%%%%%%%%%%%%%%%%%%%%%%%%%%%%%%%%%%%%%%%%%%%%%%%%%%%%%%%%%%%%%

We need to obtain $I_n(a)$ of the each order solution to calculate the higher order power spectrum. This can be obtained from Eqs.(\ref{Deq})-(\ref{Fbeq}) by using the proper initial conditions. One can rewrite the above equations by using the scale factor $a$,
\ba \fr{d^2 D}{d a^2} + \fr{3}{2a} \Bigl( 1 - w \Omega_{\DE} \Bigr) \fr{d D}{d a} - \fr{3 \Omega_{\m}}{2 a^2} D &=& 0 \label{Deqa} \\
\fr{d^2 E}{d a^2} + \fr{3}{2a} \Bigl( 1 - w \Omega_{\DE} \Bigr) \fr{d E}{d a} - \fr{3 \Omega_{\m}}{2 a^2} E &=& - \fr{3 \Omega_{\m}}{2 a^2} D^2 \label{Eeqa} \\
\fr{d^2 F_{a}}{d a^2} + \fr{3}{2a} \Bigl( 1 - w \Omega_{\DE} \Bigr) \fr{d F_{a}}{d a} - \fr{3 \Omega_{\m}}{2 a^2} F_{a} &=& - \fr{3 \Omega_{\m}}{a^2} D^3 \label{Faeqa} \\
\fr{d^2 F_{b}}{d a^2} + \fr{3}{2a} \Bigl( 1 - w \Omega_{\DE} \Bigr) \fr{d F_{b}}{d a} - \fr{3 \Omega_{\m}}{2 a^2} F_{b} &=& - \fr{3 \Omega_{\m}}{a^2} D (E - D^2) \label{Fbeqa} \ea
One can obtain the fastest growing mode solution of each order by using the proper initial condition. At early epoch, the background evolution should be identical to EdS Universe ($\Omega_{\m} = 1$) and the linear growing mode solution should be proportional to the scale factor and thus the initial conditions become $D_{g}(a_i) = a_i$ and $\fr{d D}{d a} |_{a=a_i} = 1$. Also, we assume that initial Gaussianity for the higher order solutions. It means that higher solutions should be zero at early epoch. From these, one can obtain the proper EdS fastest growing mode solutions for higher order ($E_g$, $F_{ag}$, and $F_{bg}$),
\ba D_{g}^{(\EdS)}(a_i) &=& a_i, \,\,\,\,\, \fr{d D_{g}^{(\EdS)}}{da} \Bigl|_{a=a_{i}} = 1 \, , \label{DgEdS} \\
E_{g}^{(\EdS)}(a_i) &=& -\fr{3}{7} a^2 + \fr{3}{7} a_i a = 0, \,\,\,\,\, \fr{d E_{g}^{(\EdS)}}{da} \Bigl|_{a=a_{i}} = -\fr{3}{7} a_i \, , \label{EgEdS} \\
F_{a g}^{(\EdS)}(a_i) &=& -\fr{1}{3} a^3 + \fr{1}{3} a_i^2 a = 0, \,\,\,\,\, \fr{d E_{g}^{(\EdS)}}{da} \Bigl|_{a=a_{i}} = -\fr{2}{3} a_i^2 \, , \label{FagEdS} \\
F_{b g}^{(\EdS)}(a_i) &=& \fr{70}{147} a^3 - \fr{54}{147} a_i a^2 - \fr{16}{147} a_i^2 a = 0, \,\,\,\,\, \fr{d E_{g}^{(\EdS)}}{da} \Bigl|_{a=a_{i}} = \fr{86}{147} a_i^2 \, , \label{FbgEdS} \ea

%%%%%%%%%%%%%%%%%%%%%%%%%%%%%%%%%%%%%%%%%%%%%%
\begin{figure}
\centering
\vspace{1.5cm}
\begin{tabular}{ccc}
\epsfig{file=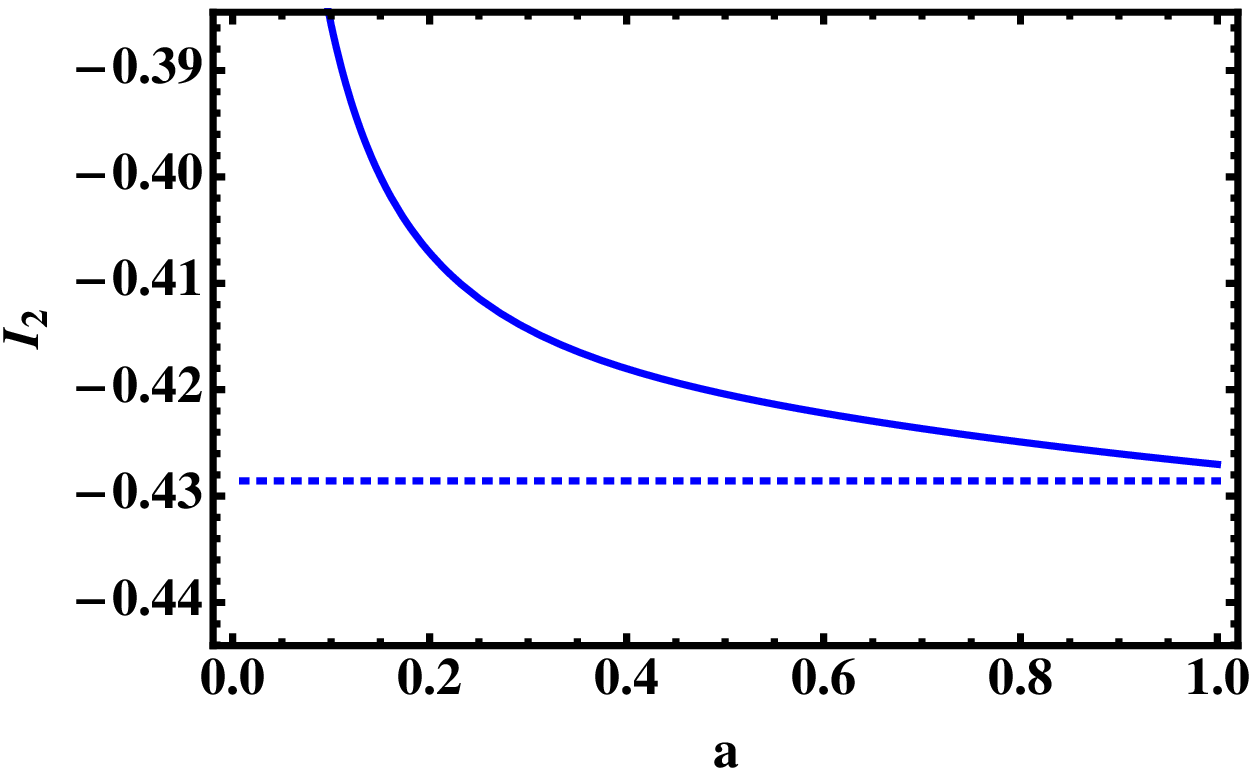,width=0.33\linewidth,clip=} &
\epsfig{file=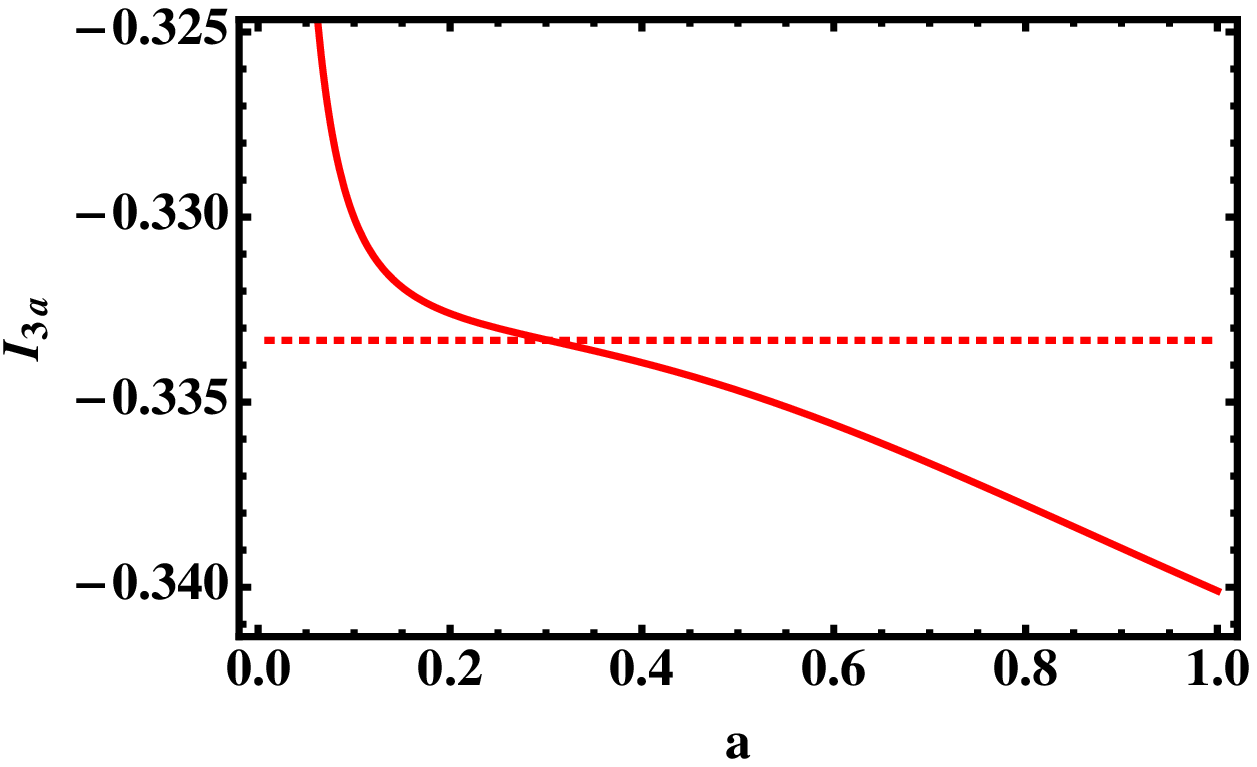,width=0.33\linewidth,clip=} &
\epsfig{file=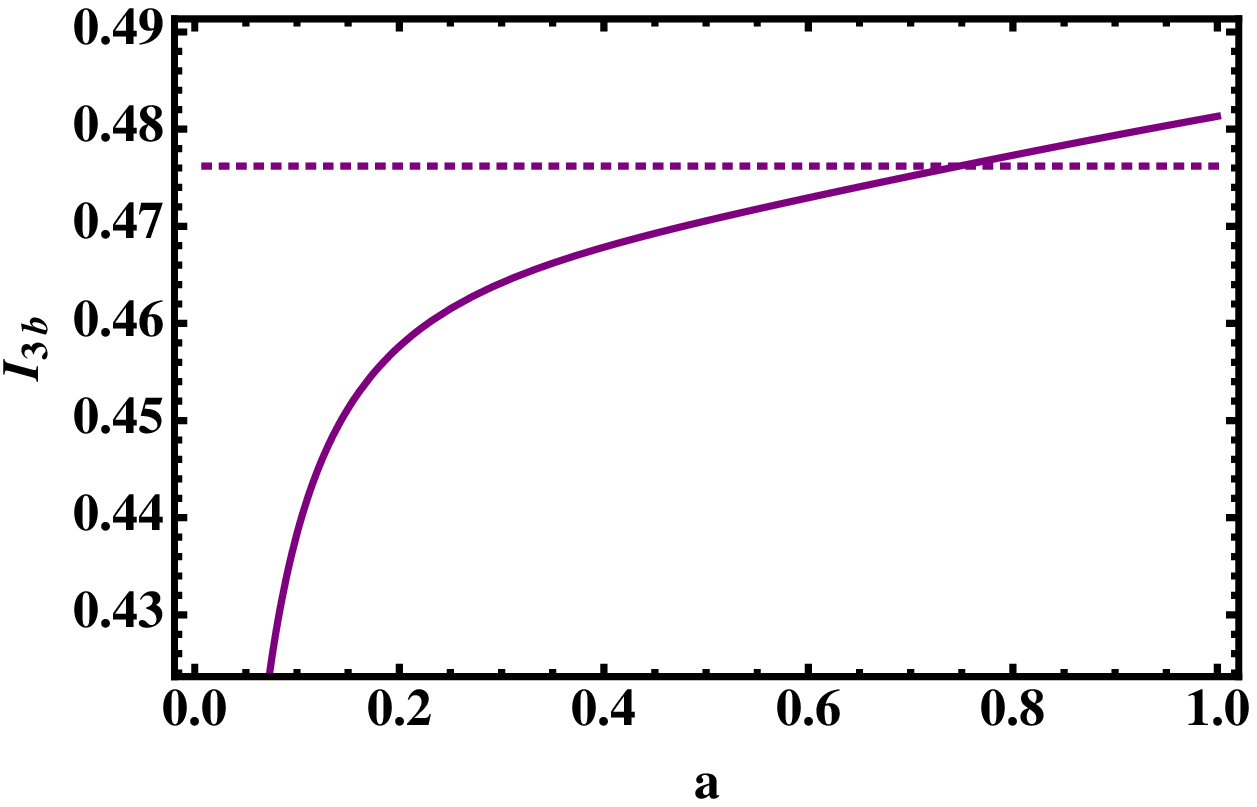,width=0.33\linewidth,clip=} \\
\end{tabular}
\vspace{-0.5cm}
\caption{The coefficients of $E$, $F_{a}$, and $F_b$ as a function of time. The dotted lines are those of EdS approximation.} \label{fig3}
\end{figure}
%%%%%%%%%%%%%%

From the above initial conditions Eqs.(\ref{DgEdS})-(\ref{FbgEdS}), one can find the higher order fastest growing mode solution for the general dark energy model and one can obtain $I_{n}(a)$ from the relation,
\be I_{2}(a) = \fr{E}{D^2}, \,\,\, I_{3a}(a) = \fr{F_{a}}{D^3}, \,\,\, I_{3b}(a) = \fr{F_{b}}{D^3} \, , \label{Is} \ee

We show the time evolutions of $I_{n}$ in Fig. \ref{fig3}. In the first panel, we show the behavior of $I_2$. As time increases, $I_2$ approaches to that of EdS assumed one. Even though we show the behavior of $I_{3a}$ in the second panel, this term does not contribute to the 1-loop power spectrum as we show. $I_{3b}$ increases as $a$ does. This is shown in the last panel of Fig. \ref{fig3}.

\end{document}